# Method for Customizable Automated Tagging

Addressing the problem of over-tagging and under-tagging text documents

Maharshi R. Pandya*
*University of North Carolina*
Chapel Hill, NC, USA
`maharshi@live.unc.edu`

Jessica Reyes*
*Kennesaw State University*
Kennesaw, GA, USA
`jreyes34@students.kennesaw.edu`

Bob Vanderheyden
*MD&I (IBM)*
Brookshaven, GA, USA
`rvanderh@us.ibm.com`

## ABSTRACT

Using author provided tags to predict tags for a new document often results in the overgeneration of tags. In the case where the author doesn't provide any tags, our documents face the severe under-tagging issue. In this paper, we present a method to generate a universal set of tags that can be applied widely to a large document corpus. Using the IBM Watson's NLU service, first, we collect keywords/phrases that we call "complex document tags" from 8,854 popular reports in the corpus. We apply LDA model over these complex document tags to generate a set of 765 unique "simple tags". In applying the tags to a corpus of documents, we run each document through the IBM Watson NLU and apply appropriate simple tags. Using only 765 simple tags, our method allows us to tag 87,397 out of 88,583 total documents in the corpus with at least one tag. About 92.1% of the 87,397 documents are also determined to be sufficiently-tagged. In the end, we discuss the performance of our method and its limitations.

## 1. Introduction

Automated keyword extraction is of utmost importance in structuring large corpus of text documents. Keywords or tags can vastly improve information retrieval (IR) tasks by giving faster and better search results [1]. They can also be used in content-based recommendation systems to provide suggestions of personalized content to the users [2].

While the use-cases of tags are enormous in the field of text mining, automatic tag extraction still remains a very challenging problem. One potential solution to this problem relies on developing machine learning models based on supervised methodologies [3]. However, the supervised methods often require an entire corpus of documents to be pre-tagged by humans. Even if there are enough resources to tag every document manually, such an effort comes with limitations such as variations in vocabulary and grammatical syntax used by human taggers to tag each document. In their work, Turney *et al.* (2000) outline repetition of candidate phrases as a potential limitation to their approach [3]. Repetition of the same keywords often give rise to the issue of over-tagging and also limit the degree to which new tags that can be discovered from the document corpus.

In recent years, we've seen the rise of hashtags on numerous social media platforms. Authors of blogs, news articles, and research papers are often required to submit keywords that best describe their documents. This solution eliminates the need of a single human tagger to tag each document since the documents are already tagged [4]. However, such approaches often give a very poor control over the variety of tags that get applied to text documents. Variations in the vocabulary and grammatical syntax used by authors to express the same tag can often result in the over-tagging of documents [3, 4]. For example, "quantum computing", "quantum computers", and "advanced quantum computers" are all keywords related to "quantum", however,

---

* Work done as IBM interns.   1

obvious variations often lead to overgeneration of tags for a single document.

For a large corpus of text documents, it is also almost inevitable to come across a document that has no tags or very few tags provided by the author. This leads to another problem of having severely under-tagged text documents. Without a tagged corpus, information retrieval tasks and recommendation platforms can face severe challenges. In this paper, we propose a simple method to address the problem of over-tagging and under-tagging of the text documents.

We utilize the IBM Watson's Natural Language Understanding (NLU) service along with the Latent Dirichlet Allocation (LDA) model to generate a set of simple monogram tags. Using the IBM Watson NLU service, we make sure every document is associated with their keywords, addressing the problem of severely under-tagged documents. Using the LDA over NLU generated keywords, we address the problem of over-tagging. Ultimately, the final set of machine-generated corpus tags will be popular across the corpus while also being general enough to tag majority of the documents in the corpus. Using the IBM Watson NLU service and the set of simple tags, we also present a simple tag application method to apply a subset of the simple tags to a new document that enters the corpus.

## 2. Related Work

Owing to its significant applications, automatic keyword extraction has been an exciting topic of research in the fields of text mining, information retrieval, and NLP. Researchers tested a diverse set of approaches to extract keywords from a large document corpus. The simplest approach utilizes document-corpus specific intrinsic statistics such as word-frequency and *tf\*idf*. Others have used graph-based methods to extract keywords. Many researchers have also tested the implementation of supervised and unsupervised methods for keyword or topic detection.

The challenge of automatic keyword extraction has been addressed in the past via simple statistical methods such as word-frequency and *tf\*idf*. These approaches simply look at word-document statistics to easily generate document-specific keywords. The *tf\*idf* approach calculates two important statistics, *tf* and *idf*, term frequency and inverse document frequency, respectively. The term frequency refers to the number of times a term appears in a document and inverse document frequency refers to the number of documents with a specific term. There are several term-weighting approaches as outlined by Salton *et al.* (1988) that can be used to determine the *tf\*idf* score for each term [5]. The sample combination of these two statistics, as shown in Equation 1, can be used for term weighting and determining importance for each of the candidate keyword in the text.

$$w_{i,j} = tf_{i,j} \times \log\left(\frac{N}{df_i}\right) \qquad (1)$$

where, $w_{i,j}$ is the weight for term $i$ in document $j$, $tf_{i,j}$ is the frequency of term $i$ in document $j$, $df_i$ is the number of documents with term $i$, and N represents the total number of documents in the corpus.

While this approach is simple, when keywords from all of the documents are combined to create a set of corpus tags for the application of corpus tags to a new document, there can be a lot of redundant tags. The *tf\*idf* statistic is ideal in matching terms to a query but it lacks ability for recognizing synonyms or any other term associations [6, 7]. This approach also requires significant text-cleaning to make sure stopwords don't end up as the keywords. When the document corpus is large, text cleaning can itself be a huge resource-consuming task, leave alone the task of extracting keywords.

Implementing supervised methods for various classification algorithms can also be helpful in obtaining machine-generated tags. In their study, Lopez *et al.* (2010), used a training set of 300 research papers from ACM and NUS to build Decision trees, Multi-Layer Perceptron (MLP) and Support Vector Machines (SVM) [8]. Using supervised methods requires a labeled dataset containing the document text and respective tags pertaining to each document. As described in the introduction, these models are less robust and severely prone to variations in human vocabulary that are inherently present in the training set [3]. These supervised classifiers also fail in discovering new tags that might be of importance in the future after the original model was trained.

Large amounts of the text documents on the internet as well as many institution-specific documents on the intranet are significantly under-tagged and, in many cases, completely missing tags. Using unsupervised methodologies, models based on these documents can be made even if they don't have tags associated with them. Graph-based approaches such as TextRank, proposed by Mihalcea and Tarau (2004), have been widely popular as an unsupervised method for keyword extraction [9]. In Graph-based methods, each node represents a keyword from a document and each edge connects related keywords. The nodes or keywords can be ranked by calculating vertex scores similar to Google's PageRank [10]. Using this approach, however, only looks at the co-occurrence relationship and ignores the semantic relationship between two nodes/keywords.

While keeping in mind the strengths and weaknesses of the aforementioned algorithms, we propose a method for keyword extraction that utilizes IBM Watson NLU for a fast, state-of-the-art extraction of keywords that recognizes semantic relationships between keywords. Using LDA, a topic modeling method, we get superior coverage of related topics, similar to TextRank, while also removing redundancy among these keywords.

## 3. Tag Extraction

The document corpus for this project contained market research reports with the author provided title, summary and content. In total, we had a total of 88,583 reports in the corpus. The pipeline for text processing normally involves pre-processing of the text by removing stop words, stemming, lemmatization, etc. Since most of the reports in our corpus contained more than 50,000 words, carrying out such pre-processing would require significant resources. Even if the text pre-processing is performed successfully, there would still be a need to build state-of-the-art keywords extraction models from scratch. Therefore, we decided to begin with extracting corpus tags from 8854 highly viewed reports, assuming these reports to be a well representation of the corpus.

The problem of text pre-processing, as well as keyword extraction, was addressed via utilizing IBM Watson NLU's outstanding capabilities. While the IBM Watson NLU service includes multiple features for text analysis such as 'entities', 'keywords', 'semantic analyzer', we decided to use 'keywords' to be the most appropriate feature for tag extraction. Using IBM Watson NLU, we were able to retrieve 'keywords' or tags in seconds for each raw document. We were also able to control the confidence scores for each tag to make sure only high-quality tags are associated with each document. For this project, tags with a confidence score of greater than 0.5 were associated with the documents.

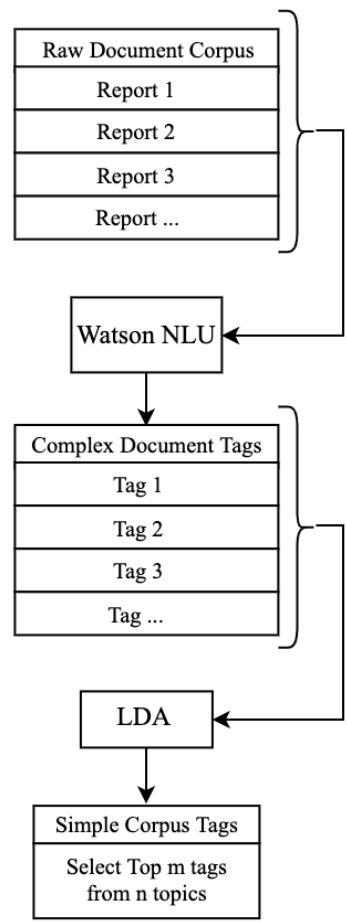

Figure 1. Tag Extraction Process

Running each document in the corpus through the IBM Watson NLU addressed the problem of limited resources for text pre-processing and extraction of quality keywords. At the same time, more importantly, it also addressed the problem of documents that were seriously under-tagged or even missing author provided tags. In fact, we completely discarded the original tags that were associated with a few of the documents in our corpus. Some of these documents were seriously over-tagged, often with completely non-relevant tags. For example, a document on Quantum Research was tagged with Financial Accounting. As a result, treating each document in the corpus as an untagged document and using IBM Watson NLU for keyword extraction provided both consistency and quality in terms of the tags that were given by the IBM Watson NLU.

The tags determined by IBM Watson NLU were highly representative of the document itself. However, the project goal was to determine a set of simple tags, not the document-specific tags, a subset of which can be applied to a wide range of documents. At large, we still didn't have control over the complex tags that resulted from the IBM Watson NLU since they weren't general enough to be used in search engine or a recommendation system. Using document specific tags is also the root cause of the over-tagging problem.

To address the problem of over-tagging and come up with a universal set of generalized corpus tags, we decided to use Latent Dirichlet Allocation (LDA) over the set of complex tags from IBM Watson NLU. The LDA, proposed by Blei *et al.* (2003), is a generative probabilistic model in which each item of a collection is modeled as a finite mixture over an underlying set of topics [11]. In practice, LDA has been widely used as an unsupervised topic modeling technique. However, in our application, the per-document output of complex tags from IBM Watson NLU can be thought of as a document itself containing only high importance terms. Using the LDA model over a set of document-specific complex tags from IBM Watson NLU allows us to generate unigrams of generalized simple tags that fall under a pre-specified number of topics.

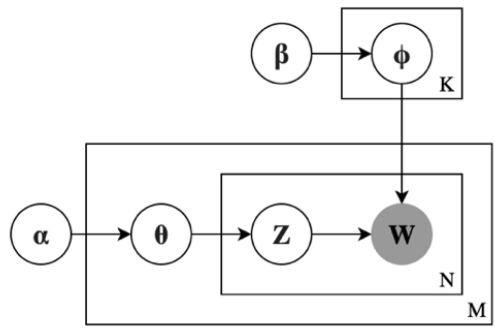

Figure 2. LDA graphical model with Dirichlet-distributed topic-keyword distributions.

In Figure 2, *M* denotes the number of documents, *N* denotes the number of words in a document, K denotes the number of topics, $\alpha$ denotes the Dirichlet-prior parameter on per-document topic distributions, $\beta$ denotes the Dirichlet-prior parameter on per-topic keyword distributions, $\theta_i$ denotes the topic distribution for i-th document, $\varphi_k$ denotes keyword distribution for k-th topic, $Z_{ij}$ denotes topic for keyword j in i-th document, and W denotes the keyword.

The sparse Dirichlet-prior parameter present in the LDA model assumes that documents in our corpus only contain few topics and each of these topics contains only a small set of words that co-occur. Words such as 'the' that might receive roughly equivalent probability scores across all topics can be easily removed. Unlike TextRank, which only looks at the term co-occurrence relationship within a document, LDA uses term co-occurrence to determine related words within a topic. Words falling under one topic may appear under other topics, however, usually with different probabilities. This allows for words within topics to be similar while between topics to be different. Taking advantage of this approach, we can selectively look at the word-topic matrix, $\varphi$, and sort keywords from largest to smallest probability values.

Our approach utilizes the LDA model, as portrayed in Figure 2, to determine the top *m* keywords that fall under each of the *K* topics. Using industry-standard 30 topics, we successfully collected top 30 keywords from each of the topics. These keywords are simple tags. Since some of these simple tags were repeated across topics, after the removal of duplicates, a final set of 765 total simple tags was extracted. The final set of 765 simple tags can be used to tag every document in the corpus via application method described in the next section.

## 4. Tag Application

Once the set of 765 simple tags was extracted, the next step in our project was to attach a subset of these corpus tags to all the documents in the corpus. Remember the 8854 highly viewed reports were only used for tag extraction and they still need to be attached with corresponding simple tags. The simple tagging process showed in Figure 3 was used for applying tags to all the documents. The same process can also be used for attaching tags to a new document that enters the corpus.

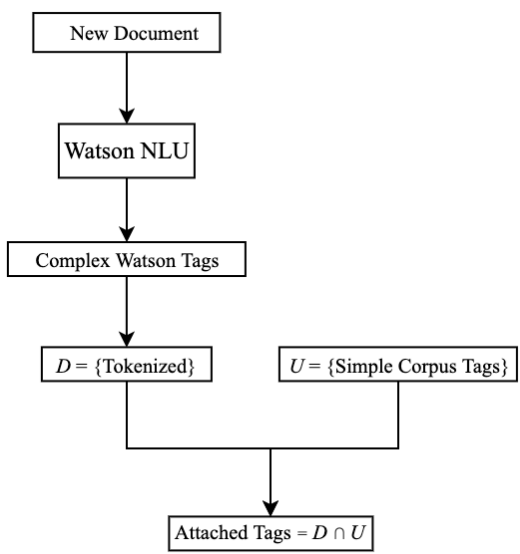

Figure 3. Tag Application Process

## 5. Results

The corpus tag extraction process, outlined in section 3, resulted in a set of 765 monogram simple tags. These tags were extracted from 8854 highly viewed documents that were assumed to be a well-representation of the entire corpus. Using this set of simple tags, each of the report in the corpus was passed through the tag application pipeline described in section 4.

Out of the total of 88,583 reports present in the corpus, exactly 87,397 reports were tagged with at least one of the simple tags. Upon inspection, 1186 reports with missing tags were discovered to be in non-English languages. Since IBM Watson NLU can extract keywords from non-English documents, they successfully went through the NLU portion in Figure 3, however, no match between the simple tags and tokenized complex tags were found, resulting in no tags being attached.

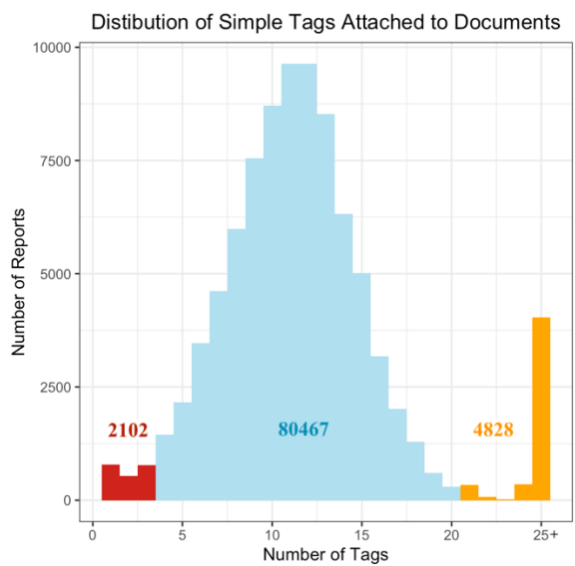

Figure 4. Frequency of reports with number of corresponding simple tags attached. Red portion indicates under-tagged reports, blue portion indicates sufficiently-tagged and orange portion indicates over-tagged reports.

Among the 87,397 reports with at least one tag attached to them, as seen in Figure 4, the majority of the reports were attached with 11-12 tags. The range for the number of tags per document was from 1 to 72. Meaning, there were few reports with 72 simple tags applied to them. For simplification, reports with more than 25 tags were combined and represented as 25+ in Figure 4. For further analysis, we made the following three classifications:
1. Reports with ≤ 3 tags are under-tagged
2. Reports with > 20 tags are over-tagged
3. Reports with > 3 and ≤ 20 tags are sufficiently-tagged.

Using the above classifications, about 2.4% of the 87,397 reports were determined to be under-tagged. On the other hand, about 5.5% of the reports were determined to be over-tagged. With 92.1% reports in the corpus determined to be sufficiently-tagged, our tag extraction and application method successfully addressed the problem of under-tagging and over-tagging.

## 6. Discussion

As a precursor to a large-scale experiment of building a recommendation engine for our platform at IBM, we were tasked with a challenge to develop a tagging process that addresses the problem of over-tagging and under-tagging present in many automatic keyword extraction methods. In this paper, we look at the details of the project and offer the following insights.

First, we found IBM Watson NLU to be an outstanding resource without which the project wouldn't have been completed in a limited time-frame. Besides saving us time and significant computing resources, IBM Watson NLU also provided us a consistent state-of-the-art keyword extraction. Since IBM Watson NLU keyword extraction also utilizes text semantics, we were able to retrieve keywords that weren't simply a result of frequency-based approaches such as *tf\*idf*. The confidence score settings in the IBM Watson NLU also give us greater control in terms of the quality of complex tags that can be considered for the LDA model.

Second, the inclusion of LDA model in our method addressed the problem of over-tagging by only extracting terms with larger probabilities in the word-topic matrix, $\varphi$, as simple tags. This method also gives us more control over the number of simple tags that we consider appropriate for extraction. Varying the number of topics in the LDA model as well as the choice of *m* in top *m* words to choose from each of the *K* topics, makes this method a greatly customizable tagging process.

Our results indicated a successful tagging process with 92.1% of the documents in the corpus to be sufficiently-tagged. The confidence score setting in IBM Watson NLU and the number of topics in the LDA model could be varied to potentially help increase this percentage. While this method addresses the problem of over-tagging and under-tagging, it requires the LDA model to be re-run on a timely basis to keep up with evolving topics that are prevalent in the market research reports.

## 7. Acknowledgements

We thank Bob Vanderheyden for providing us with an inspirational mentorship. We thank Jaspreet Singh for providing us with an expert system knowledge, whenever we were presented with a difficult situation. We thank Samuel Chow for helping us with several technical challenges. We thank entire Bluemine team for motivating us to pursue this project. We especially thank Cynthia Wang, Mahendran Nagarajan and Avinash Kohirkar for creating this opportunity and supporting us throughout our experience at IBM.